\documentclass[3p,times,procedia]{elsarticle}
\usepackage{nupha_ecrc}

\volume{00}

\firstpage{1}

\journalname{Nuclear Physics A}

\runauth{}

\jid{nupha}

\jnltitlelogo{Nuclear Physics A}

\usepackage{amssymb}

\usepackage[figuresright]{rotating}

\usepackage{graphicx}
\usepackage{dcolumn}
\usepackage{bm}
\usepackage{hyperref}
\usepackage{epsfig}
\usepackage{epstopdf} 

\newcommand{\bc}{\begin{center}}
\newcommand{\ec}{\end{center}} 
\newcommand{\be}{\begin{equation}}
\newcommand{\ee}{\end{equation}}

\newcommand{\bq}{\mathbf{q}}

\begin{document}

\begin{frontmatter}



\dochead{}

\title{Long wavelength perfect fluidity from short distance jet transport in quark-gluon plasmas}

\author[CU]{Jiechen Xu}
\author[IU,BNL]{Jinfeng Liao}
\author[CU]{Miklos Gyulassy}

\address[CU]{Department of Physics, Columbia University, 538 West 120th Street, New York, NY, USA}
\address[IU]{Physics Department and CEEM, Indiana University, 2401 North Milo B. Sampson Lane, Bloomington, IN 47408, USA}
\address[BNL]{RIKEN BNL Research Center, Building 510A, Brookhaven National Laboratory, Upton, NY 11973, USA}



\begin{abstract}
We build a new phenomenological framework that bridges the long wavelength bulk viscous transport properties of the strongly-coupled quark-gluon plasma (sQGP) and short distance hard jet transport properties in the QGP. The full nonperturbative chromo-electric (E) and chromo-magnetic (M) structure of the near ``perfect fluid'' like sQGP in the critical transition region are integrated into a semi-Quark-Gluon-Monopole Plasma (sQGMP) model lattice-compatibly and implemented into the new CUJET3.0 jet quenching framework. All observables computed from CUJET3.0 are found to be consistent with available data at RHIC and LHC simultaneously. A quantitative connection between the shear viscosity and jet transport parameter is rigorously established within this framework. We deduce the $T=160-600$ MeV dependence of the QGP's $\eta/s$: its near vanishing value in the near $T_c$ regime is determined by the composition of E and M charges, it increases as $T$ rises, and its high $T$ limit is fixed by color screening scales. 
\end{abstract}


\end{frontmatter}


\section{Introduction}
\label{sec:intro}

To probe the fundamental properties of hot quark matter and the mechanism of color confinement through ultrarelativistic nucleus-nucleus collisions, it is necessary to consider both the perturbative and nonperturbative aspects of QCD carefully in heavy-ion phenomenology. 
Present quantitative analyses of the strongly-coupled quark-gluon plasma (sQGP) created in A+A reactions at RHIC and LHC \cite{Gyulassy:2004zy} nevertheless divide in the two aspects: on the one hand, in the ``soft'' nonperturbative regime, the low transverse momentum ($p_T$) long wavelength ``perfect fluidity'' of the sQGP is described by relativistic hydrodynamical simulations; on the other hand, in the ``hard'' regime, high $p_T$ short distance jet transport properties in the QGP computed from perturbative QCD (pQCD) models are compatible with a wide range of data \cite{RHIC&LHC}. A unified framework incorporating both aspects is however missing; it is therefore challenging to translate conveniently between heavy-ion and confinement physics. 

Concentrated on pQCD, to build up such a framework, both the long and short distance transport properties of the QGP must be accounted for more systematically. In the ``soft'' sector, the ``perfect fluid'' like sQGP has a near vanishing shear viscosity to entropy density ratio $\eta/s=1/4\pi$ bounded by quantum fluctuations \cite{Danielewicz:1984ww, Kovtun:2004de}, however from leading order (LO) pQCD estimate, the QGP in the weakly-coupled limit (wQGP) has an ${\eta}/{s}\approx {0.071}({\alpha_s^2 \log (1/\alpha_s)})^{-1}$ that approaches $1$ \cite{Hirano:2005wx}. In the ``hard'' sector, it has been found that most jet energy loss models can describe the high $p_T$ light hadrons' and open heavy flavors' nuclear modification factor ($R_{AA}$) data, but the azimuthal elliptic anisotropy ($v_2$) is underestimated by $50\%$ at RHIC and LHC near-universally \cite{highpTv2}.

The above necessitates (1) exploring the full nonperturbative chromo-electric (E) and chromo-magnetic (M) structure of QCD in the region near the critical transition temperature ($T_c$), (2) developing a microscopic, lattice-compatible description of the sQGP, and (3) implementing it into a systematic pQCD jet energy loss model and testing with high $p_T$ data. The new CUJET3.0 framework achieved all of them \cite{CUJET3.0}.

\section{The CUJET3.0 framework}

\begin{figure}[!b]
\bc
\includegraphics[width=0.3225\textwidth]{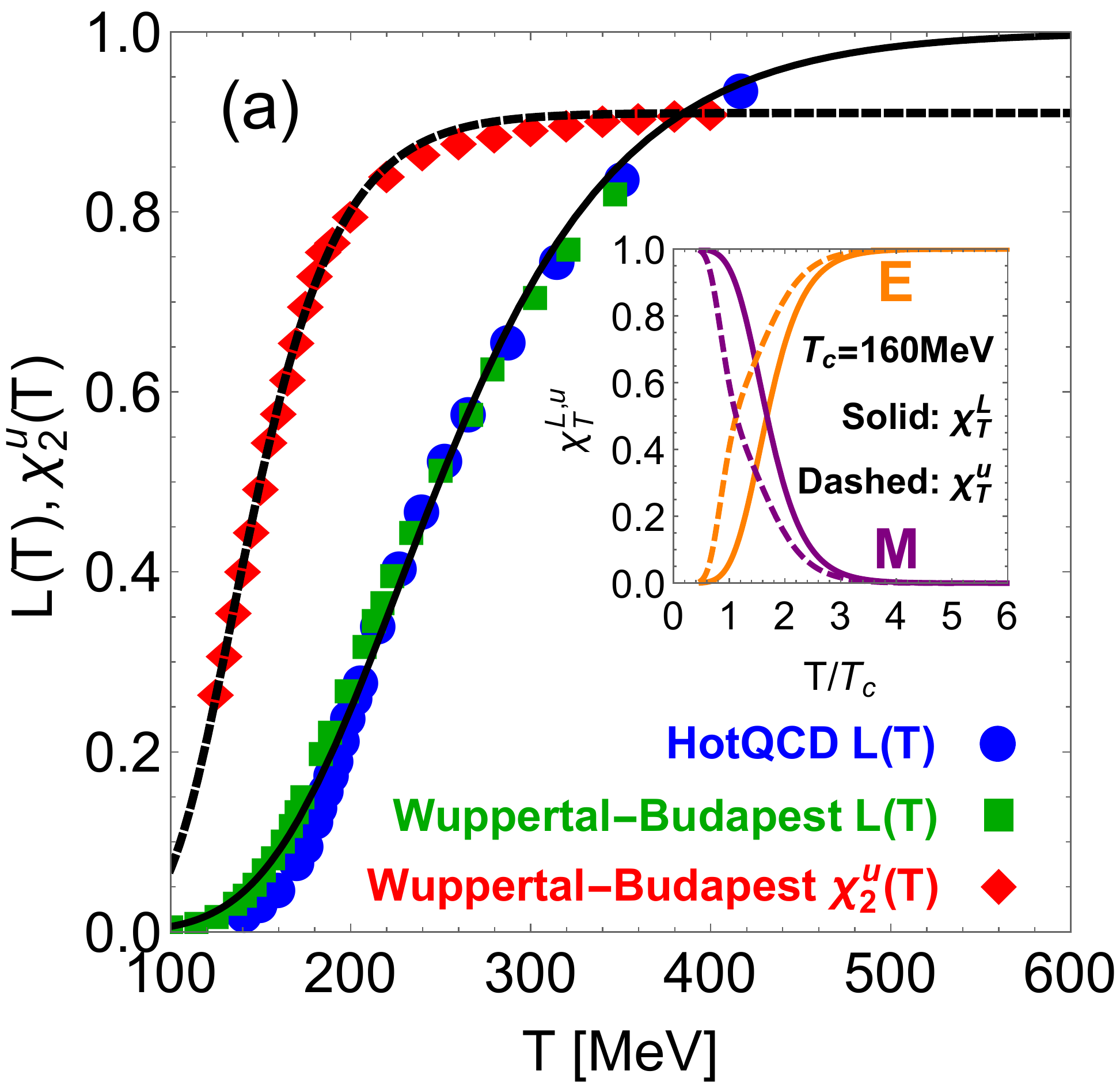}
\includegraphics[width=0.295\textwidth]{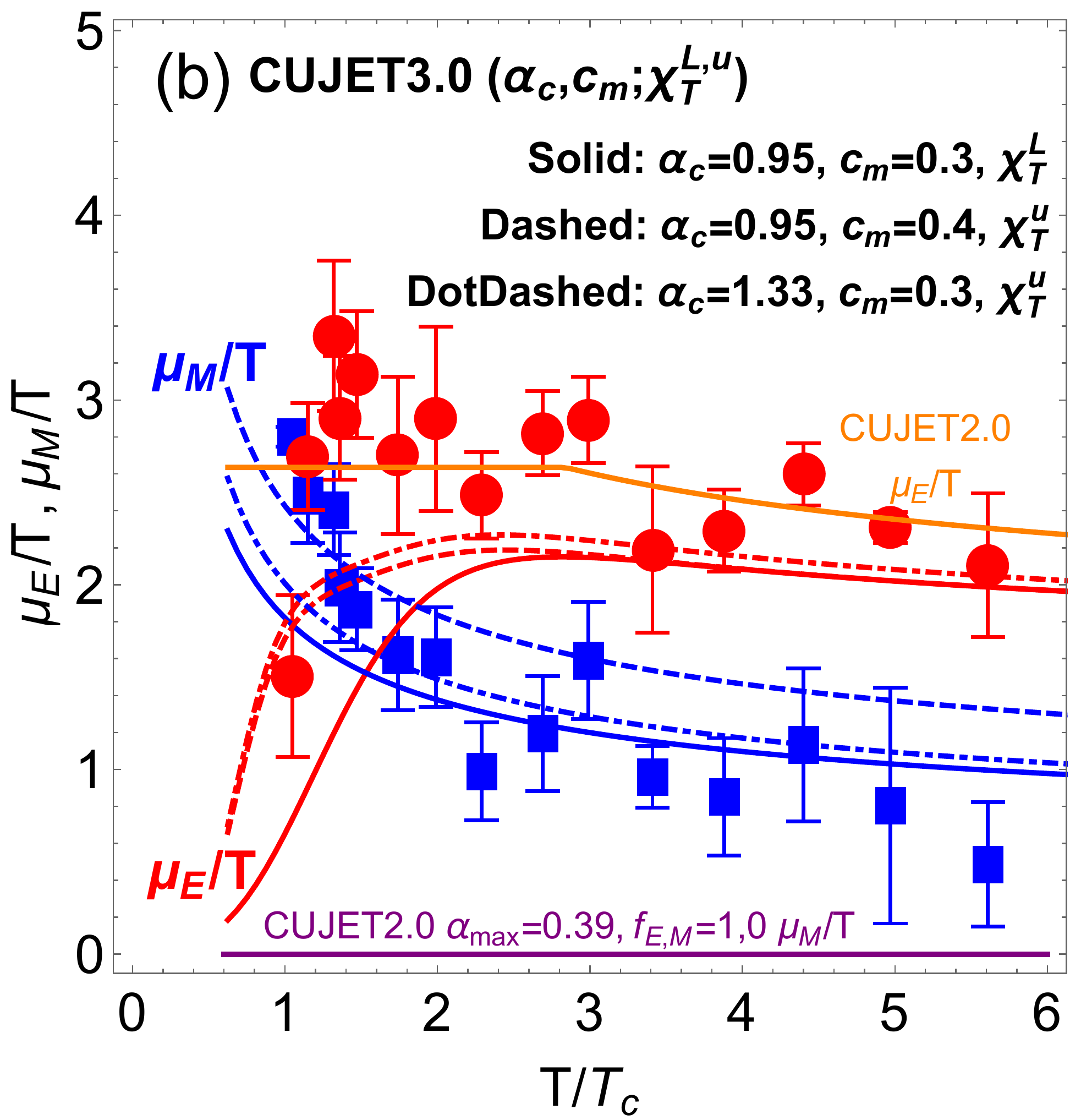}
\includegraphics[width=0.31\textwidth]{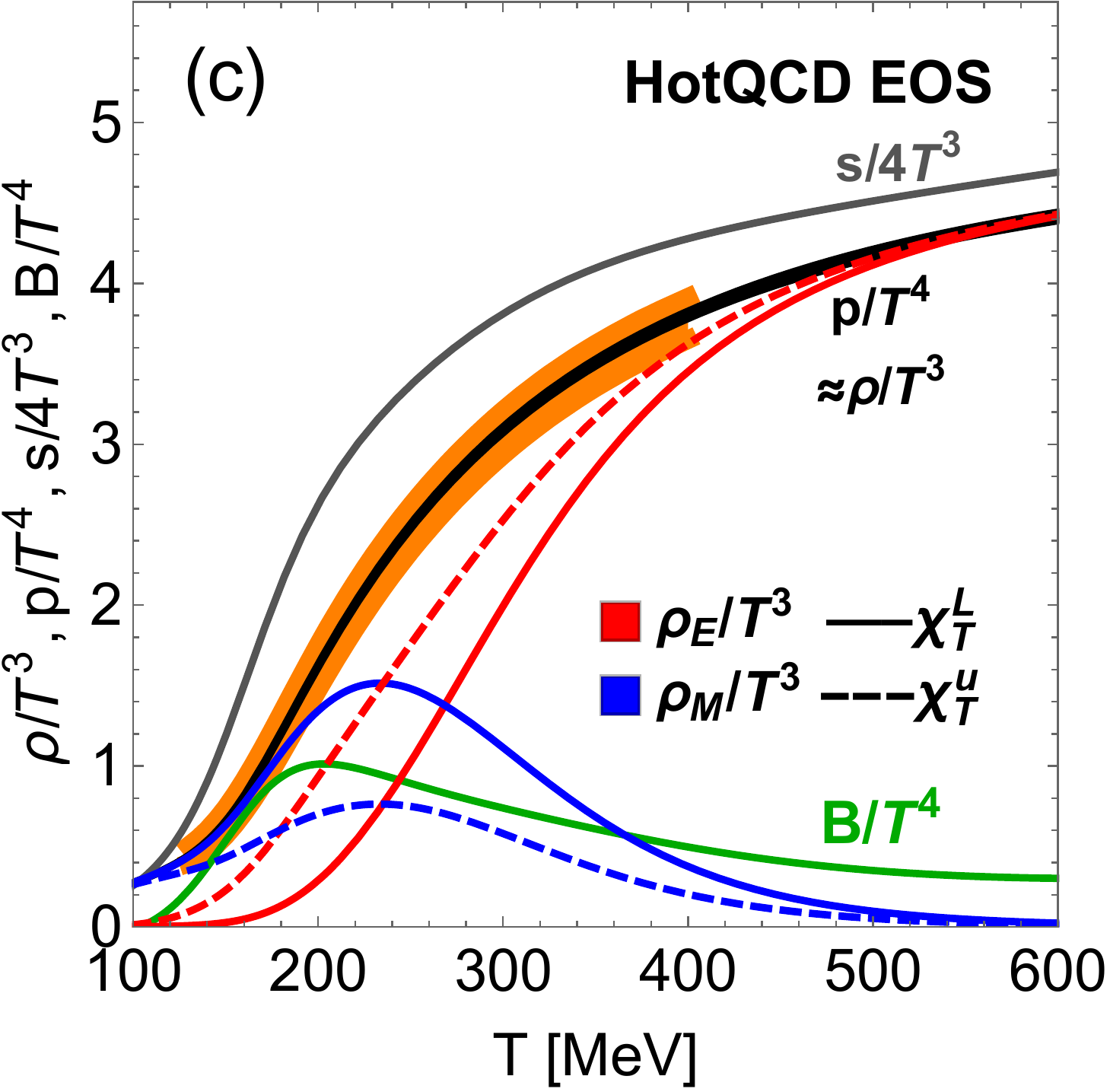}
\caption{\label{fig:lattice}
(Color online)
(a) The parameterized fit to lattice QCD data \cite{Lattice} of the renormalized Polykov loop $L$ and diagonal light quark susceptibility $\chi_2^u$ in the $\chi_T^{L}$ and $\chi_T^{u}$ scheme within CUJET3.0. The inset shows the chromo-electric $(E)$ and chromo-magnetic $(M)$ quasi-particle fractions in corresponding schemes. (b) The temperature dependence of the $E$ and $M$ screening mass $\mu_{E,M}$ in CUJET2.0 (HTL QGP) and CUJET3.0 (sQGMP) compare with lattice simulations \cite{Nakamura:2003pu}. (c) The HotQCD equation of state (EOS, pressure $p$, entropy density $s$) \cite{Lattice}, the ``bag'' pressure $(B)$, as well as the $E$ and $M$ quanta number density $\rho_{E,M}$ embedded in the CUJET3.0 framework.
}
\ec
\end{figure}

In CUJET3.0 \cite{CUJET3.0}, accounting for both chromo-electric $(E)$ and chromo-magnetic $(M)$ quasi-particles (QPs) as in the EM seesaw scenario proposed by Liao and Shuryak \cite{LS}, the dynamical running coupling DGLV \cite{DGLV} energy loss kernel in CUJET2.0 \cite{CUJET2.0} is generalized to:
\begin{eqnarray}
x\frac{dN}{dx}\propto \int d^2{q} \left[  \frac{\rho \, \alpha_s^2(\bq_\perp^2)\, f_E^2}{\bq_\perp^2 (\bq_\perp^2 + f_E^2 \mu^2)} \right]  ... \rightarrow \int d^2{q} \left[ \frac{\rho_E \left(\alpha_s(\bq_\perp^2)\alpha_s(\bq_\perp^2)\right) f_E^2}{\bq_\perp^2 (\bq_\perp^2 + f_E^2 \mu^2)} + \frac{ \rho_M \left(\alpha_E(\bq_\perp^2)\alpha_M(\bq_\perp^2)\right)  f_M^2}{\bq_\perp^2 (\bq_\perp^2 + f_M^2 \mu^2)}\;\right] ...\;\;.
\label{EMPotential}
\end{eqnarray}
Here $\alpha_s(Q^2)\equiv\alpha_E(Q^2)={\alpha_c}/[{1+\frac{9\alpha_c}{4\pi} \log(\frac{Q^2}{T_c^2})\cdot\mathrm{1}_{Q>T_c}}]$, $T_c=160\;{\rm MeV}$, and $\alpha_E\cdot\alpha_M=1$ for any $Q^2$ because of Dirac quantization \cite{LS}. The total quasi-particle number density $\rho$ consists of EQPs with fraction $\chi_T = \rho_E / \rho$ and MQPs with fraction $1-\chi_T=\rho_M/\rho$. The color electric charges are suppressed near $T_c$ as in the semi-QGP model \cite{semiQGP}, $\chi_T \equiv \chi_T^L = c_q  L + c_g L^2$, where the Polyakov loop $L(T) \propto \langle tr \mathcal{P} \exp\lbrace ig\int_{0}^{1/T} A_0 d\tau \rbrace  \rangle $ is renormalized such that $L(T\rightarrow \infty)=1$, $c_{q}$ and $c_{g}$ are Stefan-Boltzmann fraction coefficients. In the critical transition region, the semi-QGP degrees of freedom (DOFs) and emergent chromo-magnetic monopoles form a semi-Quark-Gluon-Monopole Plasma (sQGMP) \cite{CUJET3.0,OtherQGMP}. The parameter $f_E$ and $f_M$ is defined via $f_E\equiv\mu_E/\mu=\sqrt{\chi_T}$ and $f_M\equiv\mu_M/\mu=c_m g$, where $\mu_E$ and $\mu_M$ is the E and M screening mass respectively, and $g=\sqrt{4\pi\alpha_s(\mu^2)}={\mu}/({T\sqrt{1+N_f/6}})$.

The $L(T)$, $\mu_{E,M}(T)$, $\rho/T^3\sim p/T^4=\frac{1}{VT^3}\log Z$, and equation of state (EOS) are all constrained by lattice QCD data, as shown in Fig.~\ref{fig:lattice}. A theoretical uncertainty in CUJET3.0 is originated from choosing the diagonal u-quark number susceptibility $\chi_{2}^{u}(T)=\frac{\partial^{2}{(p/T^4)}}{\partial(\mu_u/T)^2}= \frac{1}{VT^3}\langle N_{u}^2\rangle$ over the Polyakov loop for the quark deconfinement rate, i.e. $\chi_T \rightarrow \chi_T^u = c_q  \chi_{2}^{u}(T)/\chi_{2}^{u}(\infty) + c_g L^2$, which will be analyzed lately. All other computational details in CUJET3.0 are the same as in CUJET2.0, including the 2+1D viscous hydrodynamical background profiles generated from VISHNU simulations \cite{VISH}.

\section{Results and discussions}
\label{result}

\begin{figure}[!t]
\bc
\includegraphics[width=0.4\textwidth]{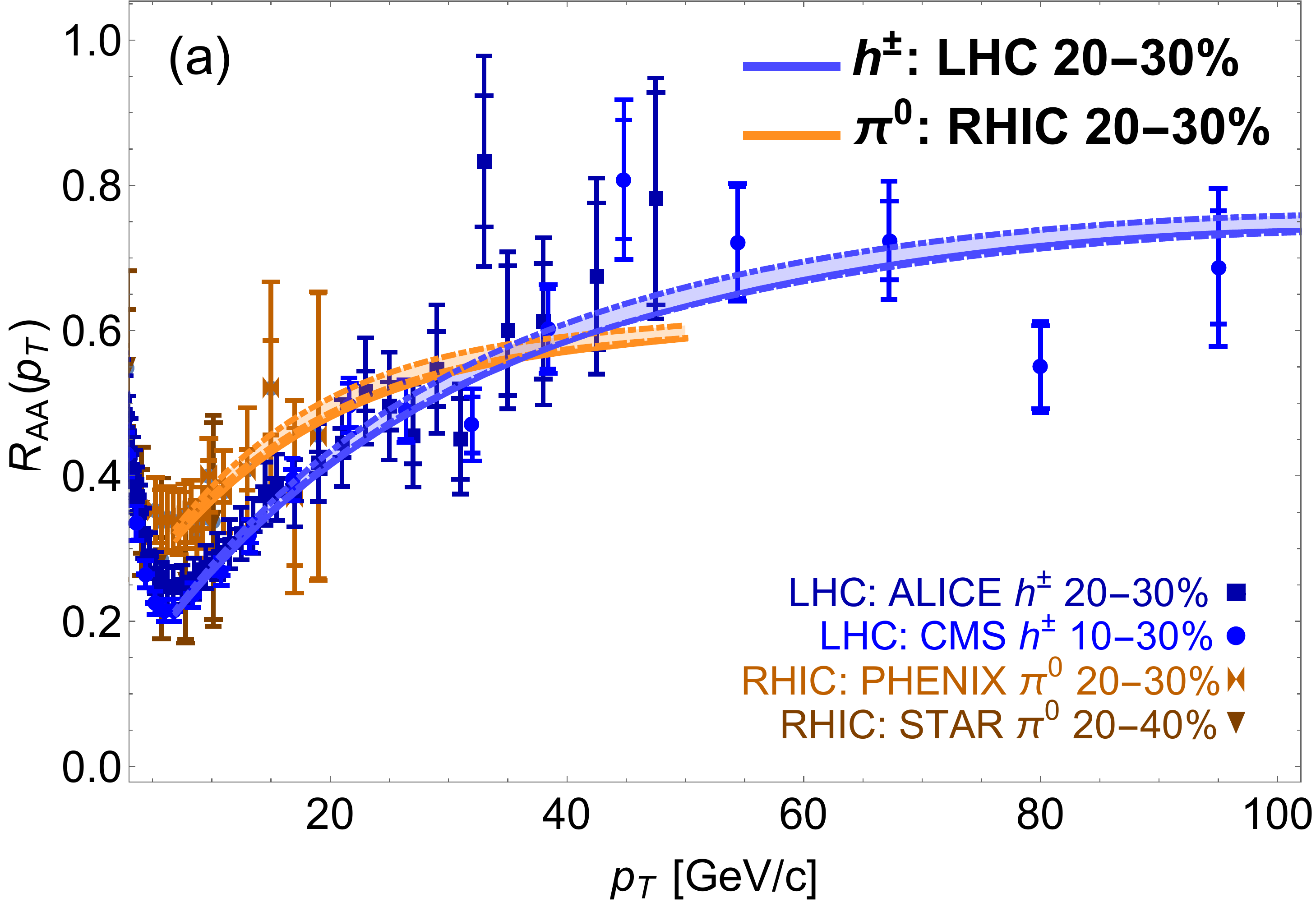}
\includegraphics[width=0.4\textwidth]{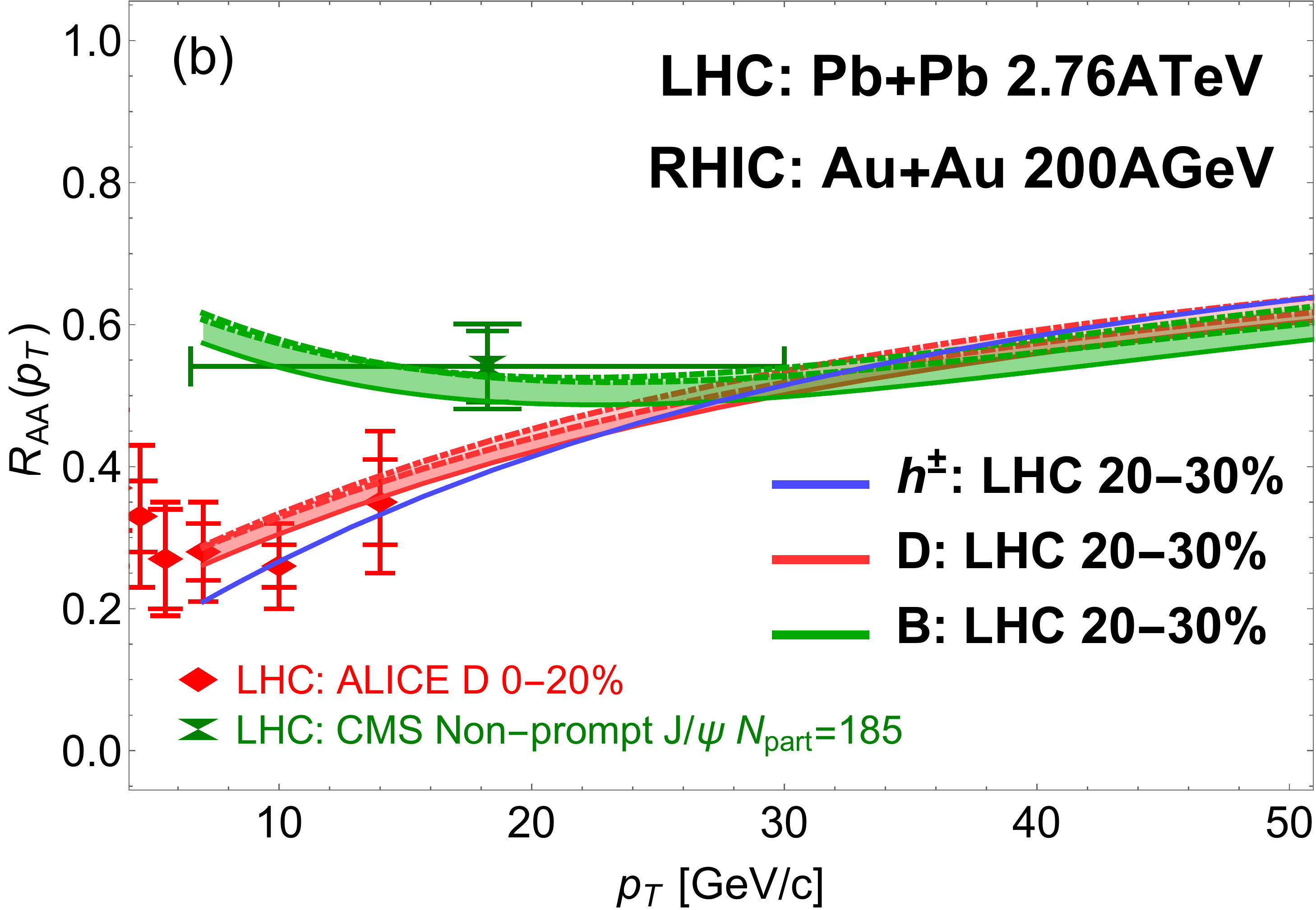}
\includegraphics[width=0.4\textwidth]{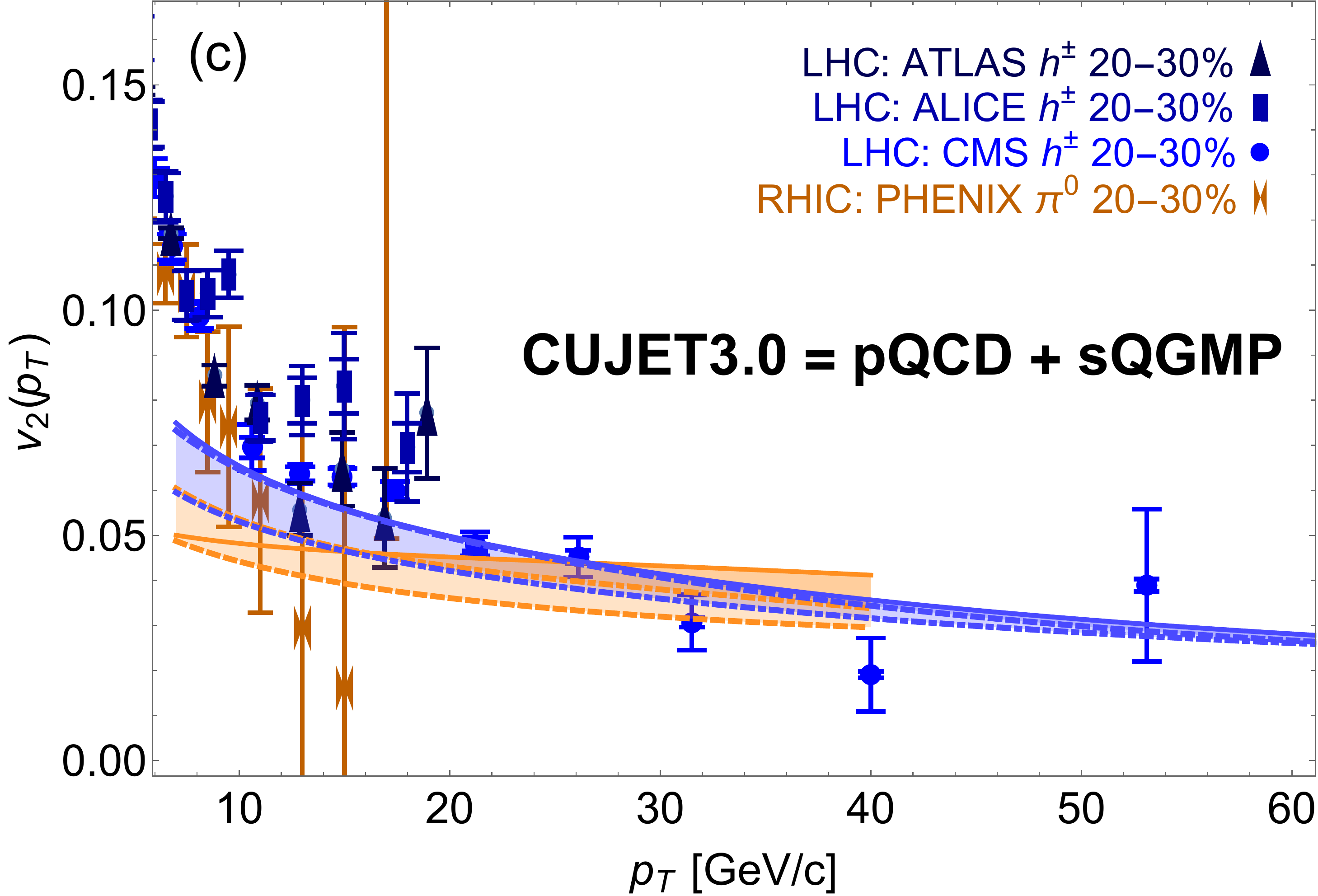}
\includegraphics[width=0.4\textwidth]{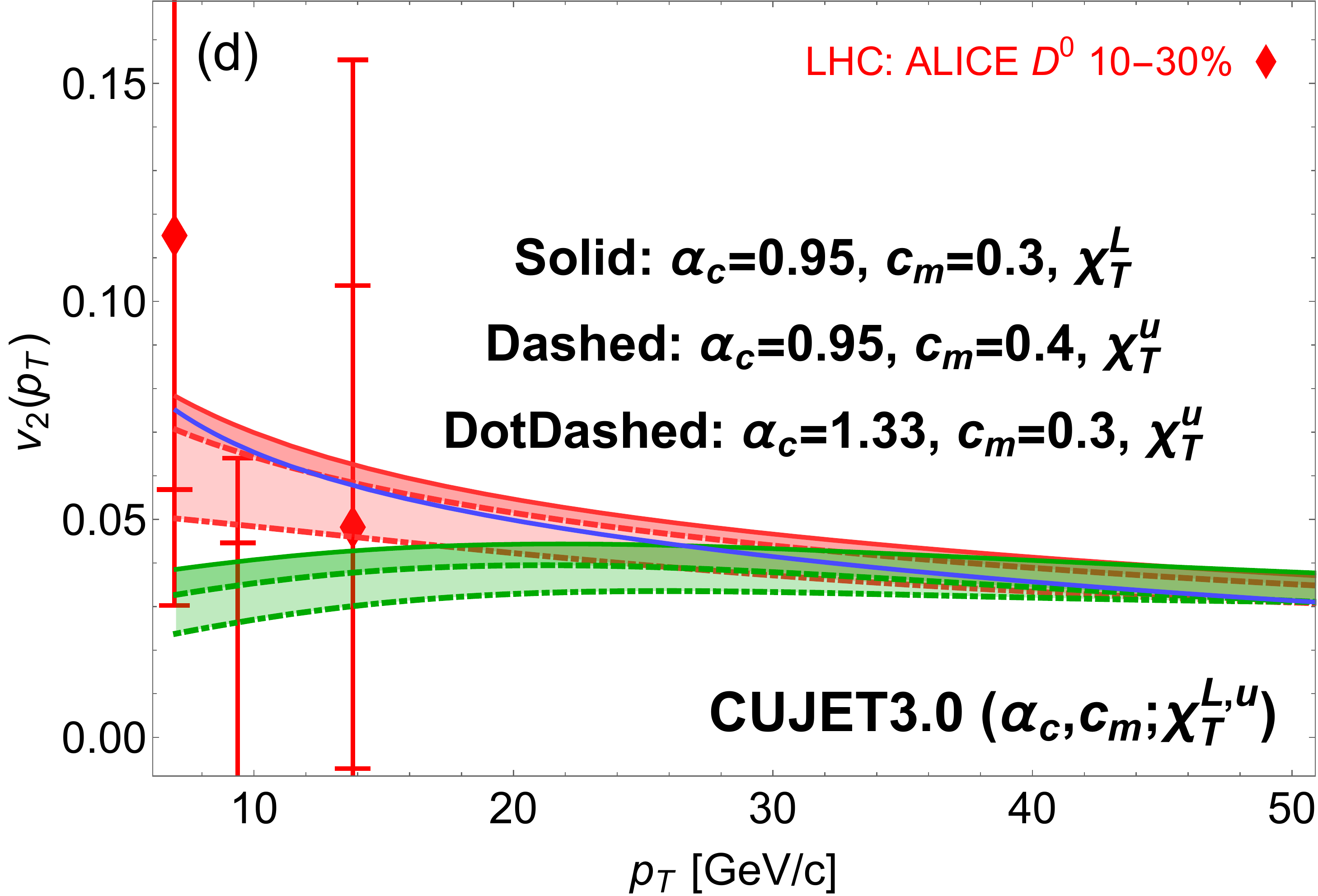}
\caption{\label{fig:obs}
(Color online)
CUJET3.0 results of (a) light hadron (LH, neutral pion $\pi^0$ and charge particle $h^\pm$)'s $R_{AA}$, (b) open heavy flavor (HF, $B$ meson and prompt $D$ meson)'s $R_{AA}$, (c) LH's $v_{2}$, and (d) HF's $v_{2}$, at high $p_T>8{\rm GeV}$ in semi-peripheral A+A collisions, compared with data from RHIC and LHC \cite{RHIC&LHC}.
The variations of predicted jet quenching observables from different schemes within CUJET3.0 suggest that data on high $p_T$ leading hadron $R_{AA}$ and $v_{2}$ in heavy-ion collisions can rigorously constrain the nonperturbative chromo-electric and chromo-magnetic structure of the QCD matter near $T_c$, and provide critical information about color confinement.
}
\ec
\end{figure}

Jet quenching observables from three different schemes in the CUJET3.0 framework will be studied: (i) $\alpha_c$=0.95, $c_m$=0.3, $\chi_T^L$; (ii) $\alpha_c$=0.95, $c_m$=0.4, $\chi_T^u$; (iii) $\alpha_c$=1.33, $c_m$=0.3, $\chi_T^u$. The parameter set $(\alpha_c,c_m)$ is constrained by the reference datum at LHC 20-30\% Pb+Pb $\sqrt{s_{NN}}=2.76$TeV  $R_{AA}^{h^\pm}(p_T=12.5{\rm GeV})\approx0.3$ and lattice date of $\mu_{E,M}(T)$ as shown in Fig.~\ref{fig:lattice}(b). Fig.~\ref{fig:obs} compares the CUJET3.0 results of leading light hadron (LH) and open heavy flavor (HF)'s $R_{AA}(p_T>8{\rm GeV})$ and $v_{2}(p_T>8{\rm GeV})$ at RHIC and LHC semi-peripheral A+A collisions with corresponding data.

For high $p_T$ LHs, all three schemes can simultaneously describe the $R_{AA}$ and $v_{2}$ data at RHIC and LHC.
The phenomenon that scheme (i) and (ii) generate a relatively larger $v_2$ than scheme (iii)
implies that the azimuthal asymmetry is sensitive to how the relative value of $\mu_E$ and $\mu_M$ inverses near $T_c$ -- the higher the inversion temperature, the longer the path length that jets interact with the monopole dominated medium at later time of the QGP evolution, the larger the high $p_T$ $v_2$.

For open heavy flavors, scheme (ii) and (iii)'s $R_{AA}$ overlap, both are larger than scheme (i)'s. Since the former two have the same color deconfinement scheme $\chi_T^u$ that is different from the latter's $\chi_T^L$, it is implicit that the HF's high $p_T$ $R_{AA}$ in CUJET3.0 is sensitive to the rate at which electric DOFs are liberated ($r_d=d\chi_T/dT$), i.e. the detailed composition of E and M DOFs near $T_c$.
Meanwhile, Fig.~\ref{fig:obs}(d) shows that the HFs' $v_2$'s are all different in scheme (i)(ii)(iii). It is therefore fair to conclude that the open charm and beauty's $R_{AA}(p_T)$ and $v_2(p_T)$ are excellent probes of the nonperturbative E and M structure of the sQGMP $(r_d,\mu_E,\mu_M)$ near $T_c$ within CUJET3.0. 

\begin{figure}[!t]
\bc
\includegraphics[width=0.45\textwidth]{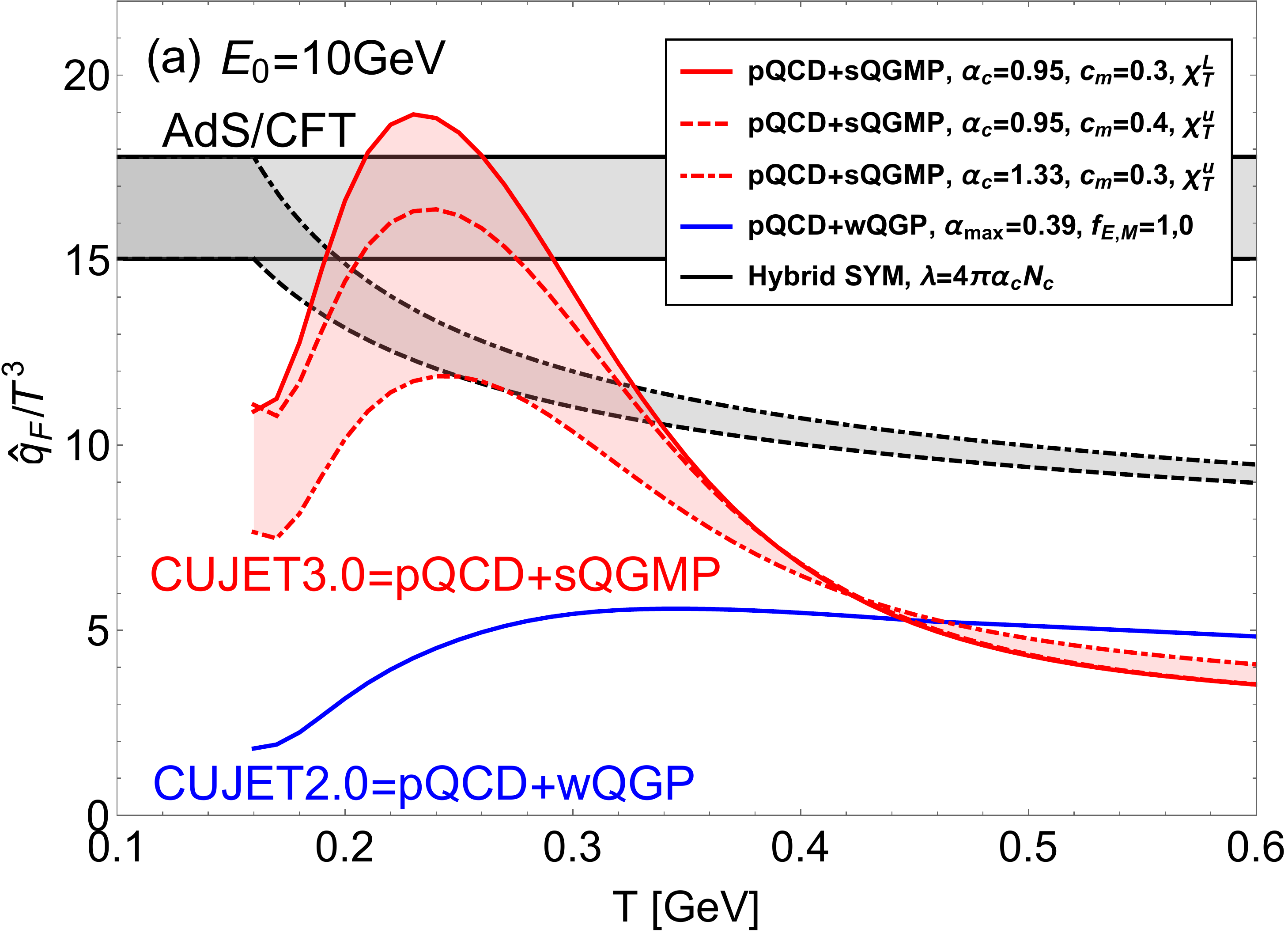}
\includegraphics[width=0.45\textwidth]{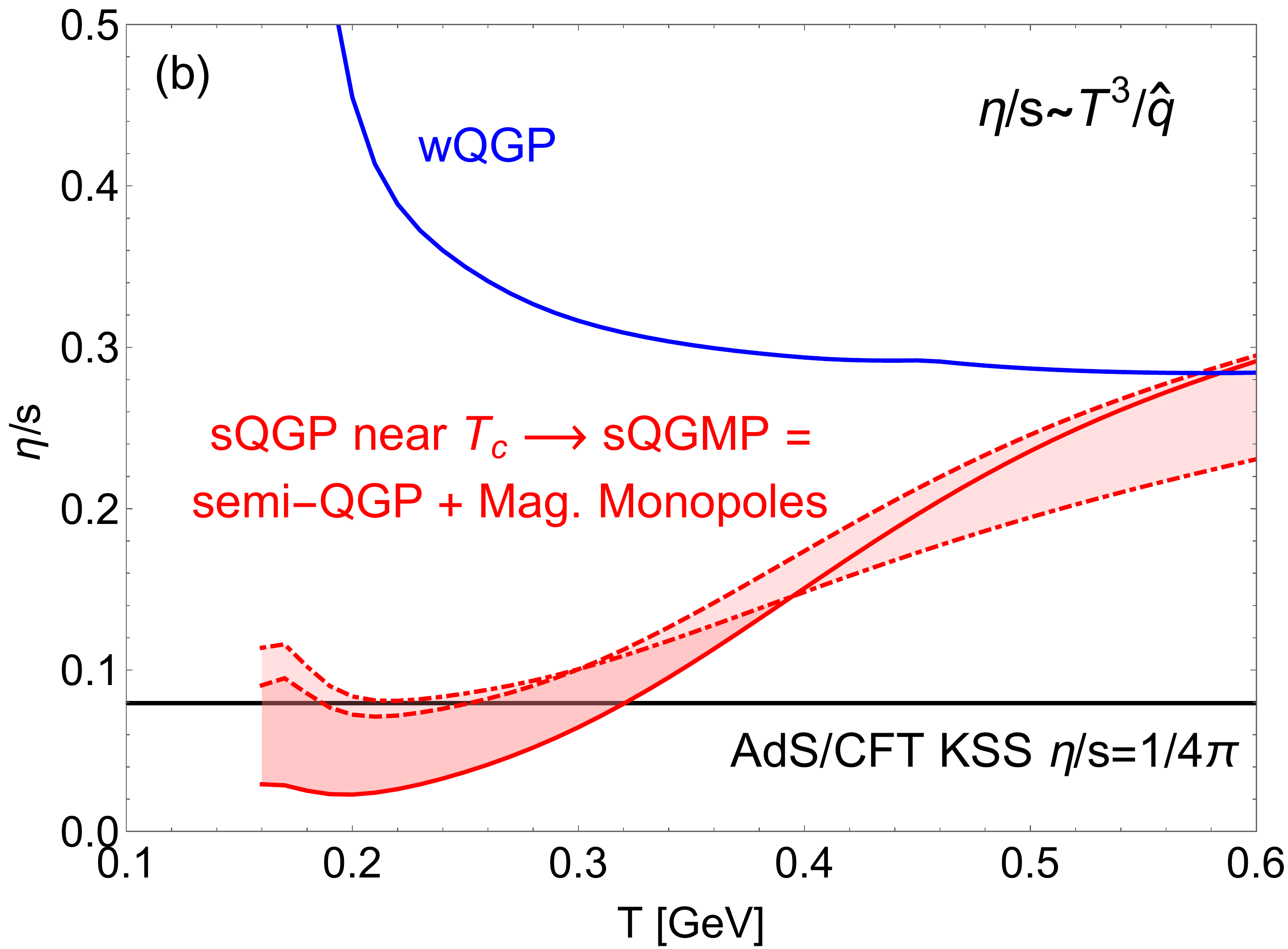}
\caption{\label{fig:qhat-etas}
(Color online)
(a) The temperature dependence of the scaled jet transport parameter $\hat{q}/T^3$ for a quark jet (in the fundamental representation $F$ of SU($N_c$=3)) with initial energy $E_0=10$ GeV in various schemes within the CUJET3.0 framework, compared with the CUJET2.0 counterpart, as well as $\mathcal{N}=4$ Supersymmetric Yang-Mills (SYM) $\hat{q}_{\rm SYM}$ results from leading order (LO) AdS/CFT calculations ($\hat{q}_{\rm SYM}=[{\pi^{3/2}\Gamma({3}/{4})}/{\Gamma({5}/{4})}]\sqrt{\lambda}T_{\rm SYM}^3$) \cite{Liu:2006ug}. Note that $3T_{\rm SYM}^3\approx T^3$ because of different number of degrees of freedom in $N_c=3$ SYM and three-flavor QCD \cite{JET}. The gray band with dashed black edges corresponds to using 't Hooft coupling $\lambda=12\pi\alpha_s(Q^2)$. (b) The shear viscosity to entropy density ratio $\eta/s$ estimated in the kinetic theory extrapolation $\eta/s\sim T^3/\hat{q}$ from jet quenching parameters in panel (a). Note that $T_c=160$ MeV. In CUJET3.0, a clear $\hat{q}_F/T^3$ maximum and $\eta/s$ minimum appear at $T\sim 1.3-1.4T_c$ where the scaled number density of emergent chromo-magnetic monopoles near $T_c$ peaks. The $(\eta/s)_{min}$ is influenced by fractions of E and M composites, hence is sensitive to confinement physics. Its value in both $\chi_T^{L,u}$ schemes converge to approximately the KSS quantum bound $\eta/s=1/4\pi$ \cite{Kovtun:2004de}. At high $T$, the $\eta/s$ from sQGMP and weakly-coupled QGP (wQGP) coincide because of similar color screening structures.
}
\ec
\end{figure}

The jet transport parameter $\hat{q}(T,E)\equiv\langle q_\perp^2 \rangle / \lambda$ in CUJET3.0 and CUJET2.0 can be extracted as in \cite{CUJET3.0} and \cite{CUJET2.0,JET} respectively. They are plotted in Fig.~\ref{fig:qhat-etas}(a). Extrapolated $\hat{q}(T,E)$ down to thermal energy scales $E\sim 3T$, one can estimate the $\eta/s$ using kinetic theory, i.e. $\eta/s =\frac{1}{s}\frac{4}{15} 
\sum_{a} \rho_a \langle p\rangle_a \lambda_a^{\perp} = \frac{18T^3}{5s}  \sum_a \rho_a/{\hat{q}}_a(T,E=3T)$, where $\rho_a(T)$ is the quasi-parton number density of type $a=q,g,m$. The $\eta/s$ results from both CUJET3.0 and CUJET2.0 are shown in Fig.~\ref{fig:qhat-etas}(b).

At high $T$ where MQPs vanish as $\chi_T\rightarrow 1$, the $\eta/s$ of sQGMP and weakly-coupled QGP (wQGP) overlap because of similar $\mu_E(T)$'s. In CUJET3.0, through fixing $\eta/s\sim T^3/\hat{q}$ at all temperatures, as $T$ cools down, $\eta/s$ drops, and a shear viscosity minimum appears at $T\sim 1.3-1.4T_c$, coinciding with the temperature where $\rho_M/T^3$ peaks as shown in Fig.~\ref{fig:lattice}(c). The value of $(\eta/s)_{min}$ is determined by the deconfinement scheme $\chi_T^{L,u}$, i.e. EQP and MQP fractions near $T_c$, and it
approaches the KSS quantum bound $\eta/s=1/4\pi$ \cite{Kovtun:2004de}. These indicate within the CUJET3.0 framework, the long wavelength ``perfect fluidity'' of the sQGP is generated from short distance jet transport properties controlled by $\hat{q}$, and a quantitative $\eta/s\sim T^3/\hat{q}$ connection is robustly established in a wide temperature range.

\section{Summary}
\label{summ}

We conclude that taking full advantage of the new CUJET3.0 jet energy loss framework, data of high $p_T$ light hadron (LH) and open heavy flavor (HF)'s $R_{AA}$ and $v_{2}$ in heavy-ion collisions at RHIC and LHC can provide stringent constraints on the nonperturbative properties of the QCD matter near $T_c$. After fixed model parameters with LH's $R_{AA}$ data, (1) LH's $v_2$ regulates the E and M screening mass difference ($\mu_{E}(T)-\mu_{M}(T)$) near $T_c$, (2) HF's $R_{AA}$ determines the rate at which color DOFs are deconfined ($r_d(T)$), (3) HF's $v_{2}$ distinguishes $r_d(T)$, $\mu_{E}(T)$ and $\mu_{M}(T)$.

In the CUJET3.0 framework, after included the semi-QGP suppression of chromo-electric charges and the emergence of chromo-magnetic monopoles in the nonperturbative near-critical QGP, the long wavelength ``perfect fluidity'' ($\eta/s\sim 1/4\pi$) is successfully generated from the short distance hard parton transport properties that are controlled by the jet quenching parameter $\hat{q}$. Within this framework, a robust $\eta/s\sim T^3/\hat{q}$ connection is established in all temperature ranges above $T_c$. Overall, CUJET3.0 provides a quantitative bridge between heavy-ion phenomenology and fundamental confinement physics.

We thank Peter Petreczky for insightful discussions. JX acknowledges helpful conversations with Gabriel Denicol, Rob Pisarski, Chun Shen and Xin-Nian Wang. The research of JX and MG is supported by U.S. DOE Nuclear Science Grants No. DE-FG02-93ER40764.
The research of JL is supported by the National Science Foundation (Grant No. PHY-1352368). 
JL also acknowledges partial support from the RIKEN BNL Research Center.





\bibliographystyle{elsarticle-num}

\begin{thebibliography}{}
\expandafter\ifx\csname url\endcsname\relax
  \def\url#1{\texttt{#1}}\fi
\expandafter\ifx\csname urlprefix\endcsname\relax\def\urlprefix{URL }\fi
\expandafter\ifx\csname href\endcsname\relax
  \def\href#1#2{#2} \def\path#1{#1}\fi

\end{thebibliography}


\begin{thebibliography}{00}

\bibitem{Gyulassy:2004zy} 
  M.~Gyulassy and L.~McLerran,
  Nucl.\ Phys.\ A {\bf 750}, 30 (2005).

\bibitem{RHIC&LHC} 
B.~Abelev {\it et al.}  [ALICE Collaboration], Phys.\ Lett.\ B {\bf719}, 18 (2013); Phys.\ Lett.\ B {\bf720}, 52 (2013); G.~Aad {\it et al.}  [ATLAS Collaboration], Phys.\ Lett.\ B {\bf707}, 330 (2012); S.~Chatrchyan {\it et al.}  [CMS Collaboration], Eur.\ Phys.\ J.\ C {\bf72}, 1945 (2012); Phys.\ Rev.\ Lett.\ {\bf109}, 022301 (2012); A.~Adare {\it et al.}  [PHENIX Collaboration], Phys.\ Rev.\ Lett.\ {\bf101}, 232301 (2008); {\bf105}, 142301 (2010); Phys.\ Rev.\ C {\bf87}, 034911 (2013); B.~I.~Abelev {\it et al.}  [STAR Collaboration], Phys.\ Rev.\ C {\bf80}, 044905 (2009).

\bibitem{Danielewicz:1984ww} 
  P.~Danielewicz and M.~Gyulassy,
  Phys.\ Rev.\ D {\bf 31}, 53 (1985).

\bibitem{Kovtun:2004de} 
  P.~Kovtun, D.~T.~Son and A.~O.~Starinets,
  Phys.\ Rev.\ Lett.\  {\bf 94}, 111601 (2005).
  
\bibitem{Hirano:2005wx} 
  T.~Hirano and M.~Gyulassy,
  Nucl.\ Phys.\ A {\bf 769}, 71 (2006).
  
\bibitem{Lattice} 
A.~Bazavov {\it et al.}, Phys.\ Rev.\ D {\bf 80}, 014504 (2009); S.~Borsanyi {\it et al.} [Wuppertal-Budapest Collaboration], JHEP {\bf 1009}, 073 (2010); S.~Borsanyi, Z.~Fodor, S.~D.~Katz, S.~Krieg, C.~Ratti and K.~Szabo, JHEP {\bf 1201}, 138 (2012).

\bibitem{Nakamura:2003pu} 
  A.~Nakamura, T.~Saito and S.~Sakai,
  Phys.\ Rev.\ D {\bf 69}, 014506 (2004).

\bibitem{highpTv2} 
  B.~B.~Abelev {\it et al.} [ALICE Collaboration],
  Phys.\ Rev.\ C {\bf 90}, no. 3, 034904 (2014);
S.~Cao, G.~Y.~Qin and S.~A.~Bass,
Phys.\ Rev.\ C {\bf 92}, 024907 (2015);
  B.~Betz and M.~Gyulassy,
  JHEP {\bf 1408}, 090 (2014)
  [JHEP {\bf 1410}, 043 (2014)];
  arXiv:1503.07671 [hep-ph];
  D.~Molnar and D.~Sun,
  arXiv:1305.1046 [nucl-th].

\bibitem{CUJET3.0} 
  J.~Xu, J.~Liao and M.~Gyulassy,
  Chin.\ Phys.\ Lett.\  {\bf 32}, 092501 (2015);
  arXiv:1508.00552 [hep-ph].

\bibitem{LS}
  J.~Liao and E.~Shuryak,
  Phys.\ Rev.\ C {\bf 75}, 054907 (2007);
  Phys.\ Rev.\ Lett.\  {\bf 101}, 162302 (2008);
    {\bf 102}, 202302 (2009).
 
\bibitem{DGLV} 
M.~Gyulassy {\it et al.}, Nucl.\ Phys.\ B {\bf 594}, 371 (2001); M.~Djordjevic {\it et al.}, Nucl.\ Phys.\ A {\bf 733}, 265 (2004); Phys.\ Rev.\ Lett.\ {\bf101}, 022302 (2008); S.~Wicks {\it et al.}, Nucl.\ Phys.\ A {\bf 784}, 426 (2007); A.~Buzzatti {\it et al.}, Phys.\ Rev.\ Lett.\ {\bf108}, 022301 (2012).


\bibitem{CUJET2.0} 
  J.~Xu, A.~Buzzatti and M.~Gyulassy,
  JHEP {\bf 1408}, 063 (2014);
    Nucl.\ Phys.\ A {\bf 932}, 128 (2014).

\bibitem{semiQGP}
  R.~D.~Pisarski,
  Phys.\ Rev.\ D {\bf 74}, 121703 (2006);
Y.~Hidaka and R.~D.~Pisarski,
	Phys.\ Rev.\ D {\bf 78}, 071501 (2008);
  {\bf 81}, 076002 (2010);
A.~Dumitru {\it et al.},
  {\bf 86}, 105017 (2012);
C.~Gale {\it et al.},
Phys.\ Rev.\ Lett.\  {\bf 114}, 072301 (2015).



\bibitem{OtherQGMP}
B.~G.~Zakharov,
JETP Lett.\  {\bf 101}, 587 (2015);
A.~Iwazaki,
arXiv:1511.02271 [hep-ph].


\bibitem{VISH} 
H.~Song and U.~W.~Heinz, Phys.\ Rev.\ C {\bf 78}, 024902 (2008);
C.~Shen, U.~Heinz, P.~Huovinen and H.~Song, Phys.\ Rev.\ C {\bf 82}, 054904 (2010);
  C.~Shen, Z.~Qiu, H.~Song, J.~Bernhard, S.~Bass and U.~Heinz,
  Comput.\ Phys.\ Commun.\  {\bf 199}, 61 (2016).

\bibitem{Liu:2006ug} 
H.~Liu, K.~Rajagopal and U.~A.~Wiedemann,
Phys.\ Rev.\ Lett.\  {\bf 97}, 182301 (2006).


%
%
%


\bibitem{JET} 
  K.~M.~Burke {\it et al.} [JET Collaboration],
    Phys.\ Rev.\ C {\bf 90}, 014909 (2014).






\end{thebibliography}








\end{document}